\documentclass[12pt]{iopart}
\usepackage{epsfig}
\usepackage{bm}
\usepackage{iopams}

\newcommand{\nn}{\nonumber}
\newcommand{\cmz}{\,^{\mbox{\tiny cm}}\!z}
\newcommand{\cmbmz}{\,^{\mbox{\tiny cm}}\!{\bm{z}}}
\newcommand{\cmdotz}{\,^{\mbox{\tiny cm}}\!\dot{z}}
 
\newcommand{\nM}{\,^{\mbox{\tiny n}}\!M}
\newcommand{\nG}{\,^{\mbox{\tiny n}}G}
\newcommand{\pnM}{\,^{\mbox{\tiny pn}}\!M}
\newcommand{\pnG}{\,^{\mbox{\tiny pn}}G}

\newcommand{\ndotG}{\,^{\mbox{\tiny n}}{\dot G}}
\newcommand{\nddotG}{\,^{\mbox{\tiny n}}{\ddot G}}

\newcommand{\ndotM}{\,^{\mbox{\tiny n}}\!{\dot M}}
\newcommand{\nddotM}{\,^{\mbox{\tiny n}}\!{\ddot M}}

\newcommand{\pndotM}{\,^{\mbox{\tiny pn}}\!{\dot M}}

\newcommand{\deltahatxl}{\delta_{(a_1a_2...} \delta_{a_{2q-1}a_{2q}} \hat{x}_{a_{2q+1}... a_l)}}

\newcommand{\be}{\begin{equation}}
\newcommand{\ee}{\end{equation}}
\newcommand{\bea}{\begin{eqnarray}}
\newcommand{\eea}{\end{eqnarray}}
\newcommand{\ve}{\varepsilon}
\DeclareMathSymbol{\R}{\mathbin}{AMSb}{"52}
\begin{document}
\jl{6}
\title{Spin and energy evolution equations for a wide class of extended bodies}
\author{\'{E}tienne Racine\footnote{email: er55@cornell.edu}}
\address{Center for Radiophysics and Space Research, Cornell University, Ithaca, New York, 14853}
\date{\today}

\begin{abstract}
We give a surface integral derivation of the leading-order evolution equations for the spin and energy of a relativistic body interacting with other bodies in the post-Newtonian expansion scheme. The bodies can be arbitrarily shaped and can be strongly self-gravitating. The effects of all mass and current multipoles are taken into account. As part of the computation one of the 2PN potentials parametrizing the metric is obtained. The formulae obtained here for spin and energy evolution coincide with those obtained by Damour, Soffel and Xu for the case of weakly self-gravitating bodies. By combining an Einstein-Infeld-Hoffman-type surface integral approach with multipolar expansions we extend the domain of validity of these evolution equations to a wide class of strongly self-gravitating bodies. This paper completes in a self-contained way a previous work by Racine and Flanagan on translational equations of motion for compact objects. 
   

\end{abstract}
\maketitle

\section{Introduction}

In a recent paper by Racine and Flanagan \cite{paperI}, henceforth referred to as paper I, explicit post-1-Newtonian equations of motion for the center-of-mass worldline of an astrophysical body interacting with other bodies have been derived from a surface integral method. These equations of motion are valid for a large class of strongly self-gravitating objects and include the effects of all mass and current multipole moments, which are defined by surface integrals of the post-Newtonian gravitational potentials.

The main purpose of this paper is to pursue the development of the program initiated in paper I in two ways. First, using a surface integral method pioneered by Einstein, Infeld and Hoffmann \cite{eih} and later used in many other references [e.g. Landau and Lifshitz \cite{landaulifshitz}, Thorne and Hartle \cite{thornehartle})] we give a self-contained derivation of leading order evolution equations for the spin [Eq.(\ref{sdot}) below] and energy [Eq.(\ref{energyevolution}) below] of a given body. These two evolution equations have been derived by Damour, Soffel and Xu \cite{dsxII} for the case of weakly self-gravitating bodies. We prove here their validity for compact objects including neutron stars and quiescent black holes. These equations were needed in paper I to obtain final expressions for the translational equations of motion and thus the present sequel article completes this computation. Second, we construct an explicit expansion of the general solution of a certain post-2-Newtonian field equation in a region between two concentric spheres, using a standard method of post-Newtonian theory\cite{bdell}. In this paper we use our solution of a particular post-2-Newtonian field in surface integral computations, as part of the derivation of the evolution equation for the energy. In addition the expansions of post-2-Newtonian fields outlined in this paper could be used in principle to give a more self-contained derivation of the equations of motion of paper I. In paper I, an argument invoking globally weak fields was used to deduce the form of the laws of motion for strong-field sources. Here we show how to rigorously derive, in principle, the laws of motion [Eq. (4.3b) of paper I or Eq.(4.21b) of \cite{dsxII}] without invoking any results obtained in the context of globally weak fields. These computations would follow the general outline of section \ref{sec:energy}, in which the energy evolution equation is derived.  


We first give our conventions below in section \ref{sec:notation}, summarize the formalism in section \ref{sec:summary}, detail in section \ref{sec:inverselaplacian} a construction of a general solution to the relevant post-2-Newtonian field equation, then derive, in section \ref{sec:spin}, the evolution equation for the spin and end with the derivation of the evolution equation for the energy in section \ref{sec:energy}.

\subsection{Notation and conventions}\label{sec:notation}

We use geometric units in which $G=c=1$.  We use
the sign conventions of Misner, Thorne and Wheeler \cite{mtw}; in
particular we use the metric signature $(-,+,+,+)$.  
Greek indices ($\mu$,$\nu$ etc...) run from 0 to 3 and denote
spacetime indices, while Roman indices ($a$, $b$, $i$,$j$, etc...) run
from 1 to 3 
and denote spatial indices.  
The spacetime coordinates will generically be denoted by $(x^0,x^i) =
(t,x^i)$.  Spatial indices are raised and lowered
using $\delta_{ij}$, and repeated spatial indices are contracted
regardless of whether they are covariant or contravariant indices.
We denote by $n^i$ the unit vector $x^i/r$, where $r = |{\bm{x}}| =
\sqrt{\delta_{ij} x^i x^j}$. 

When dealing with sequences of spatial indices, we use the
multi-index notation introduced by Thorne \cite{thorne} as modified
slightly by Damour, Soffel and Xu \cite{dsxI}.  
We use $L$ to denote the sequence of $l$ indices $a_1 a_2 \ldots a_l$,
so that for any $l$-index tensor\footnote{Here by ``tensor'' we mean an object which transforms as a tensor under the symmetry group SO(3) of the Euclidean spatial metric $\delta_{ij}$, not a spacetime tensor.} $T$ we have
\be\label{ell}
T_L \equiv T_{a_1a_2... \,a_l} .
\ee
If $l=0$, it is
understood that $T_L$ is a scalar. If $l<0$ then $T_L\equiv
0$.  We define $L-1$ to be the sequence of $l-1$ indices $a_1 a_2
\ldots a_{l-1}$, so that 
\be
T_{L-1} \equiv T_{a_1a_2... \,a_{l-1}}.
\ee
We also define 
\be\label{iell}
T_{iL} \equiv T_{ia_1a_2...a_l}.
\ee
For symmetric tensors we use the convention that the "$l$" version of (\ref{iell}) and the "$l+1$" version of (\ref{ell}) define the same tensor. For example $T_i$ and $T_{a_1}$ are defined to be exactly the same tensor. We define $K$ to be the sequence of $k$
spatial indices $b_1 b_2 \ldots b_k$, so that
\[ T_K \equiv T_{b_1b_2...\, b_k}.\]
Repeated multi-indices are subject to the Einstein summation
convention, as in $S_L T_L$.  We also use the notations
\be
x^L \equiv x^{a_1a_2...\,a_l} \equiv x^{a_1}x^{a_2}...\,x^{a_l}
\ee
and
\be
\partial_L \equiv \partial_{a_1a_2...\,a_l} \equiv
\partial_{a_1}\partial_{a_2} ...\,\partial_{a_l}. 
\ee
We use angular brackets to denote the operation of taking the symmetric
trace-free part of a tensor. Thus for any tensor $T_L$, we define
\be
T_{<L>} \equiv \mbox{STF}_L(T_L).
\ee
where $\mbox{STF}_L$ means taking the symmetric trace-free projection on the
indices $L$. For example, if $l=2$, we have  
\be
T_{<L>} = T_{<a_1a_2>} = \frac{1}{2}\left(T_{a_1a_2} + T_{a_2a_1}\right) - \frac{1}{3}\delta_{a_1a_2}T_{jj}.
\ee
The general formula for the STF projection of a tensor is

\be
T_{<L>} \equiv \sum_{k=0}^{[l/2]}c_k^l
\delta_{(a_1a_2}...\delta_{a_{2k-1}a_{2k}}S_{a_{2k+1}...a_l)j_1j_1...j_kj_k},
\ee
where
$[l/2]$ is the largest integer less than or equal to $l/2$ and $S_L$ is the symmetric part of $T_L$. The
coefficients $c^l_k$ are given by
\be
c^l_k = (-1)^k\frac{l!}{(l-2k)!}\frac{(2l-2k-1)!!}{(2l-1)!!(2k)!!}.
\ee
We also use the notation

\be
\hat{x}_L \equiv x_{<L>}.
\ee


\section{Summary of the formalism}\label{sec:summary}

The physical systems considered in this paper are systems composed of a collection of arbitrarily structured astrophysical bodies. It is assumed that the bodies are separated enough, and their internal dynamics slow enough, that the gravitational field in between the bodies is well approximated by a post-Newtonian expansion (see paper I for a detailed discussion). This assumption allows a wide class of strongly self-gravitating objects like quiescent black holes and neutron stars. It excludes however bodies emitting very strong bursts of gravitational waves. This restriction applies to strong gravitational waves produced by internal motions. Strong gravitational waves due to orbital motions are allowed. A more detailed discussion of the class of objects and systems encompassed by our analysis is given in paper I. 

The systems we consider are expected to be strong sources of gravitational radiation, potentially detectable by modern gravitational wave observatories. For example in coalescing binary systems, the waveforms of the emitted gravitational radiation are expected to carry lots of information, and full exploitation of the expected observations will require accurate theoretical models of the waveforms. One of the main motivation of the work reported in paper I and in the present sequel paper is to improve the precision of theoretical models of gravitational waveforms emitted by systems of compact objects.  

A formalism to describe such systems to post-1-Newtonian order in orbital velocities has been developed in detail in paper I. This section summarizes the important features of that formalism and serves as a simplified reading guide for sections II and III of paper I. The work reported in paper I is, in part, a generalization of the formalism devised by Damour, Soffel and Xu \cite{dsxII,dsxI,dsxIII,dsxIV} and Brumbeg and Kopeikin \cite{kop,Brumberg1989,Brumberg1991} to strongly self-gravitating objects. One of the main ingredients of the framework presented in paper I is, as in the DSX work, a collection of coordinate systems with a set of post-Newtonian field equations and solutions associated with each coordinate system. There is a global coordinate system that covers the region in between and far from all bodies. In this global coordinate system, each object is treated on an equal footing. There is also one coordinate system per body that covers a shell-like region enclosing that body. Each of these coordinate systems is adapted to the enclosed body. The precise mathematical meaning of the word "adapted" will be spelled out below. The region of overlap between an adapted coordinate system and the global coordinate system is called the buffer region of the body. All adapted coordinate systems and the global coordinate system are conformally cartesian, satisfy harmonic gauge conditions and cover vacuum regions in which the post-Newtonian expansion is assumed to be valid. In this paper however, we will only need the gravitational field expressed in adapted coordinate systems. All the computations requiring the global coordinate system needed for this paper have already been performed in paper I. 

Another important part of the formalism is the definition of a set of mass and current mulitpole moments characterizing each body and a set of gravitoelectric and gravitomagnetic tidal moments felt by each body. These moments are defined as parameters of potentials that are general solutions to vacuum post-Newtonian field equations in a particular gauge. We refer the reader to Appendix E of paper I for the explicit definitions of the moments in terms of surface integrals of the post-Newtonian gravitational potentials. It is important to note that these moments are invariant under a post-1-Newtonian transformation of the time coordinate [cf. Eq.(2.30) in paper I], which is a gauge freedom analogous to that of electromagnetism. In the DSX papers, mass and current multipole moments are instead defined in terms of compact support integrals over the matter distribution. As shown in paper I, our field-based definitions of mass and current multipole moments coincide with DSX's multipole moments in the case of weakly self-gravitating bodies. The field-based definition is more general since it can also be applied to strongly self-gravitating bodies. In that case however, the moments cannot be expressed in terms of compact support integrals over the matter distribution. 

The respective expansions of the metric $g_{\mu\nu}$ and of the Gothic metric $\mathfrak{g}^{\alpha\beta} \equiv \sqrt{g}g^{\alpha\beta}$ to the required order\footnote{See the discussion at the beginning of section IV C in paper I.}  for the computations detailed in this paper are, in the coordinate system $(s_A,y^j_A)$ adapted to body $A$ [cf. Eqs.(4.11) and (4.12) of paper I]

\bea\label{nm00}
g_{00} = - \frac{1}{\ve^2} - 2\Phi^A - 2\ve^2\big[(\Phi^A)^2 + \psi^A\big] + O(\ve^4) \\\label{nm0i}
g_{0i} = \ve^2\zeta^A_i + O(\ve^4) \\\label{nmij}
g_{ij} = \delta_{ij}\Big[1 - 2\ve^2\Phi^A + \ve^4\big(-2\psi^A + 2(\Phi^A)^2 + \chi^A_{kk}\big)\Big]  - \ve^4\chi^A_{ij} + O(\ve^5)
\eea
and

\bea\label{gm00}
\mathfrak{g}^{00} = -\ve + 4\ve^3\Phi^A - \ve^5\left[8(\Phi^A)^2 - 4\psi^A + \chi^A_{kk}\right] + O(\ve^6) \\
\mathfrak{g}^{0i} = \ve^3\zeta^A_i + O(\ve^5) \\\label{gmij}
\mathfrak{g}^{ij} = \frac{1}{\ve}\delta^{ij} + \ve^3\chi^A_{ij} + O(\ve^4). 
\eea
The terms of order $\ve^4$ in $g_{ij}$ are not required for the computation of the spin evolution equation but they will be needed later for the computation of the evolution equation for the energy. In Eqs.(\ref{nm00})-(\ref{nmij}), the parameter $\ve$ is the standard post-Newtonian bookkeeping parameter. Our generic time coordinate $s_A$ differs from usual time coordinates $\hat{t}$ by a factor of $\ve$:

\be
\hat{t} = \frac{s_A}{\ve}.
\ee
By a usual time coordinate ${\hat t}$ we mean a time coordinate with
the property that $g_{{\hat t}{\hat t}} \to -1$ as $\varepsilon \to
0$. The motivation behind this choice of a rescaled generic time coordinate is explained in section IIA of paper I. The functions appearing in Eqs.(\ref{nm00}) - (\ref{gmij}) satisfy, in vacuum, the following field equations

\be\label{fieldeqA}
\nabla^2\Phi^A = 0 ,
\ee
\be\label{fieldeqB}
\nabla^2\zeta^A_i = 0,
\ee
\be\label{fieldeqC}
\nabla^2 \psi^A = \frac{\partial^2\Phi^A}{\partial s_A^2}, 
\ee
\be\label{fieldeqD}
\nabla^2 \chi^A_{ij} = 4\frac{\partial \Phi^A}{\partial y^i_A}\frac{\partial \Phi^A}{\partial y^j_A} - 2\delta_{ij}\frac{\partial \Phi^A}{\partial y^k_A}\frac{\partial \Phi^A}{\partial y^k_A},
\ee
where

\be
\nabla^2 = \delta_{ij}\frac{\partial}{\partial y^i_A}\frac{\partial}{\partial y^j_A}.
\ee
These fields are subject to the following (harmonic) gauge conditions

\be\label{zetagauge}
4\frac{\partial \Phi^A}{\partial s_A} + \frac{\partial \zeta^A_i}{\partial y^i_A} = 0
\ee
and

\be\label{chigauge}
\frac{\partial \zeta^A_i}{\partial s_A} + \frac{\partial \chi^A_{ij}}{\partial y^j_A} = 0.
\ee
The general solution to the set of equations (\ref{fieldeqA})-(\ref{fieldeqC}) in a region of the form $0 < r_- \leq |\bm{y}_A| \leq r_+$ can be written as

\bea\label{adaptedA}
\Phi^A(y^j_A,s_A) = \sum_{l=0}^\infty
\frac{(-1)^{l+1}}{l!}\nM_L^A(s_A)\partial_L\frac{1}{|\bm{y}_A|} -
\frac{1}{l!}\nG_L^A(s_A)y_A^L ,\\\label{adaptedB}   
 \zeta_i^A(y^j_A,s_A) = \sum_{l=0}^\infty
\frac{(-1)^{l+1}}{l!}Z_{iL}^A(s_A)\partial_L\frac{1}{|\bm{y}_A|}  -
\frac{1}{l!}Y_{iL}^A(s_A)y_A^L, \\\nn
\fl \psi^A(y^j_A,s_A) = \sum_{l=0}^\infty
\left\{\frac{(-1)^{l+1}}{l!}
\left[\pnM_L^A(s_A) +  \frac{(2l+1)}{(l+1)(2l+3)}\dot{\mu}^A_L(s_A)\right]\partial_L 
\frac{1}{|\bm{y}_A|}   \right. \\ \label{adaptedC} 
\fl \left. + \frac{(-1)^{l+1}}{l!}\!\nddotM_L^A(s_A)\partial_L \frac{|\bm{y}_A|}{2} - \frac{1}{l!}
\left[\pnG_L^A(s_A) -\dot{\nu}^A_L(s_A)\right]y^L_A -
\frac{1}{l!}\frac{|\bm{y}_A|^2}{2(2l + 3)}\nddotG_L^A(s_A) y^L_A \right\}, 
\eea
where

\bea
\mu^A_L = Z^A_{jjL} \\
\nu^A_L = Y^A_{<L>}.
\eea
All tensors, or moments, parametrizing the solutions (\ref{adaptedA})-(\ref{adaptedC}) are STF on their multi-index $L$ only. The moments $Z^A_{iL}$ and $Y^A_{iL}$ can be written in terms of fully STF tensors by using the following reduction formulas \cite{paperI, dsxI}

\bea\label{Zstfreduction}
Z^A_{iL} = \frac{4}{l+1}\ndotM^A_{iL} - \frac{4l}{l+1}\epsilon_{ji<a_l}S^A_{L-1>j} + \frac{2l-1}{2l+1}\delta_{i<a_l}\mu^A_{L-1>} ,\\\label{Ystfreduction}
Y^A_{iL} = \nu_{iL} + \frac{l}{l+1}\epsilon_{ji<a_l}H^A_{L-1>j} - 4\frac{2l-1}{2l+1}\delta_{i<a_l}\ndotG^A_{L-1>} 
\eea
where overdots mean time derivative with respect to the time argument (here $s_A$). The moments $S^A_L$ and $H^A_L$ introduced above are STF on all their indices.

The moments $\nM^A_L$ and $\pnM^A_L$ are respectively the Newtonian and post-Newtonian mass multipoles of body $A$; $\nG^A_L$ and $\pnG^A_L$ are the Newtonian and post-Newtonian gravitoelectric tidal moments felt by body $A$. The moments $S^A_L$ are the current multipole moments and $H^A_L$ are the gravitomagnetic tidal moments. The moments $\nM^A_L$, $\pnM^A_L$, $\nG^A_L$ and $\pnG^A_L$ are defined for $l \geq 0$ while $S^A_L$ and $H^A_L$ are defined for $l \geq 1$. The $\mu_L$ and $\nu_L$ moments are called gauge moments and give information on the coordinate system used.

The general solution to the field equation (\ref{fieldeqD}) given the expansion (\ref{adaptedA}) for the Newtonian potential has not been derived in the literature, as far as the author knows, and is the topic of section \ref{sec:inverselaplacian}. We therefore do not discuss it here.

Now there are many harmonic coordinates systems in the region $r_- \leq |\bm{y}_A| \leq r_+$  in which the metric has the form (\ref{nm00})-(\ref{nmij}). This gauge freedom has been characterized in detail by DSX \cite{dsxI}, Kopeikin \cite{kop}, Klioner and Voinov \cite{klionervoinov} and in paper I. We define the adapted coordinate system to be the coordinate system that takes full advantage of this gauge freedom to simplify the solutions (\ref{adaptedA})-(\ref{adaptedC}) in a particular way. It has been shown in paper I that this gauge freedom makes it always possible to choose coordinates such that the following conditions hold:

\bea\label{adaptedconditionA}
\nM^A_i = 0, \\
\pnM^A_i = 0, \\\label{adaptedconditionmu}
\mu^A_L = 0 \,\, \forall \,\,l \geq 0,\\
\nG^A = 0, \\
\pnG^A = 0, \\
\nu^A_L = 0 \,\,\forall \,\, l \geq 1,\\\label{adaptedconditionB}
H^A_i = 0. 
\eea

Equations (\ref{adaptedconditionA})-(\ref{adaptedconditionB}) define the adapted coordinate system. By that we mean that after conditions (\ref{adaptedconditionA})-(\ref{adaptedconditionB}) have been imposed, there is no gauge freedom left over, except constant translations and rotations. Note however that the evolution equations we obtain in the end are valid in all mass-centered conformally cartesian harmonic gauges since the multipole and tidal moments appearing in the evolution equations are invariant under the transformation

\be
\bar{s}_A = s_A + \ve^4 \beta,
\ee
where $\beta$ is a harmonic function. Finally, let us recall here a result from paper I that will be needed later on. In paper I expressions for the tidal moments $\nG^A_L$, $\pnG^A_L$ and $Y^A_{iL}$ were derived [respectively Eqs. (5.25b), (5.30b) and (5.27b) of paper I] in terms of the center-of-mass worldlines and all the mass and current multipoles of all the other bodies $B \neq A$ part of the system. Here we will need only the expression for $\nG^A_L$ for $l \geq 2$:

\be\label{tidalG}
\nG^A_L = \sum_{B \neq A}\sum_{k=0}^\infty \frac{(-1)^k}{k!}(2k+2l-1)!!\nM^B_K \frac{n^{BA}_{<KL>}}{r_{BA}^{k+l+1}} \,\,\,\, \mbox{for} \,\, l \geq 2,
\ee
where 

\be
r_{BA} = |\cmbmz^B - \cmbmz^A|
\ee
and 

\be
\bm{n}^{BA} = (\cmbmz^B - \cmbmz^A)/r_{BA}. 
\ee
The center-of-mass worldlines $\cmbmz^A$ are defined in section V C of paper I. 
\section{Solutions of 2PN field equations}\label{sec:inverselaplacian}

The derivation of the energy evolution equation in section \ref{sec:energy} requires the general solution to (\ref{fieldeqD}). It can be written as

\be\label{chisol}
\chi_{ij} = \chi_{ij}^{\mbox{\scriptsize p}} + \sum_{l=0}^\infty\left[ \frac{(-1)^{l+1}}{l!}C_{ijL}\partial_L\frac{1}{r} - \frac{1}{l!}B_{ijL}x^L\right] ,
\ee
where $\chi_{ij}^{\mbox{\scriptsize p}}$ is a particular solution to (\ref{fieldeqD}), with $C_{ijL}$ and $B_{ijL}$ being STF on $L$.  Above we dropped all the labels $A$ and rename $(s_A,y^j_A) \rightarrow (t,x^j)$ to abbreviate the notation, since no confusion about the coordinate system can arise: we work exclusively in the coordinate system adapted to body $A$. It is very important to note that the quantities $C_{ijL}$ and $B_{ijL}$ are partially determined by the gauge condition (\ref{chigauge}). These moments can therefore depend on $\dot{Z}_{iL}$ and $\dot{Y}_{iL}$ as well as on quantities that are intrinsically of post-2-Newtonian order. A similar situation has been encountered at post-1-Newtonian order [cf. Eqs.(\ref{Zstfreduction})-(\ref{Ystfreduction})] where the moments $Z_{iL}$ and $Y_{iL}$ parametrizing the gravitomagnetic potential depend partially on the Newtonian moments $\nM_L$ and $\nG_L$, this dependence being determined by the harmonic gauge condition 

\be
4\dot{\Phi} + \partial_i\zeta_i = 0.
\ee
The particular solution $\chi_{ij}^{\mbox{\scriptsize p}}$ is constructed by solving the following elliptic equation

\be\label{elliptic}
\nabla^2 f_L = r^p x^L ,
\ee
One convenient solution is obtained by using Eq.(A21a) and (A37) of Ref.\cite{bdell}, which are respectively 

\be\label{bdone}
x^L = r^l \sum_{q=0}^{[l/2]}\frac{l!(2q-1)!!}{(2q)!(l-2q)!}\frac{(2l-4q+1)!!}{(2l-2q+1)!!}\delta_{(a_1a_2}...\delta_{a_{2q-1}a_{2q}}\hat{n}_{a_{2q+1}...a_l)},
\ee
and

\be\label{bdtwo}
\nabla^2\left[r^{\lambda}\hat{n}^L\right] = (\lambda - l)(\lambda + l +1)r^{\lambda-2} \hat{n}^L.
\ee
First one starts from the ansatz

\be\label{ellipticansatz}
f_L =  \sum_{q=0}^{[l/2]} \left[a_{q}(l,p) + b_{q}(l,p)\log \frac{r}{b}\right] r^{p+ 2q + 2} \delta_{(i_1i_2...}\delta_{i_{2q-1}i_{2q}} \hat{x}_{i_{2q+1}... i_l)},
\ee
the $\log$ terms being needed to invert (\ref{bdtwo}) in the special cases $\lambda = l$ and $\lambda = -l-1$. The constant $b$ is an arbitrary scale. Applying the Laplacian on both sides of (\ref{ellipticansatz}) and requiring that $f_L$ satisfies (\ref{elliptic}) yields



\bea\label{abqlp}
a_q(l,p) = \frac{l!}{(2q)!!(l-2q)!}\frac{(2l-4q+1)!!}{(2l-2q+1)!!}\frac{1}{(p+2q+2)(p-2q+2l+3)} \\
b_q(l,p) = 0,
\eea
for $q \neq q_c$, $q_c$ being the possible value of $q$ for which $(p+2q+2)$ or $(p-2q+2l+3)$ vanishes. Note that there exist a $q_c$ only if $p < 0$. For $q = q_c$, $a_{q_c}(l,p)$ is undetermined and is therefore an integration constant that can be absorbed in the scale $b$ of the log term given in (\ref{ellipticansatz}). We thus set $a_{q_c}(l,p)$ to zero by convention. For $q=q_c$, $b_{q_c}(l,p)$ is given by

\be
b_{q_c}(l,p) = \frac{l!}{(2q_c)!!(l-2q_c)!}\frac{(2l-4q_c+1)!!}{(2l-2q_c+1)!!}\frac{1}{(2p+2l+5)}.
\ee
To shorten equations in the rest of the paper, we introduce the following abbreviations

\be
\left(\delta \hat{x}\right)^q_L \equiv \deltahatxl
\ee
and

\be
c_q(l,p;r) = a_q(l,p) + b_q(l,p) \log \frac{r}{b}.
\ee

To solve for $\chi_{ij}$, we first write the gradient of the Newtonian potential as follows

\bea
\partial_i \Phi = \sum_{l=0}^\infty \frac{1}{l!} \frac{(2l+1)!! }{r^{2l+3}}x^{<iL>} \nM_L - \frac{1}{l!}\nG_{iL} x^L \nn \\
= \sum_{l=0}^\infty \frac{x^{<L>}}{l!} \left[l \frac{(2l-1)!!}{r^{2l+1}}\delta_{i<a_l}\nM_{L-1>} - \nG_{iL}\right],
\eea
where we used

\be\label{id}
\partial_L \frac{1}{r} = (-1)^l\frac{(2l + 1)!!}{(2l+1)} \frac{x^{<L>}}{r^{l+1}}.
\ee
We then have

\bea
\fl \partial_i \Phi \partial_j \Phi = \sum_{k=0}^\infty\sum_{l=0}^\infty \frac{x^Kx^L}{k!l!} \Bigg[kl\frac{(2k-1)!!(2l-1)!!}{r^{2k+2l+2}}\delta_{i<a_k}\nM_{K-1>} \delta_{i<a_l}\nM_{L-1>} + \nG_{iK}\nG_{jL} \nn \\ - k\frac{(2k-1)!!}{r^{2k+1}}\delta_{i<a_k}\nM_{K-1>}\nG_{jL} - l\frac{(2l-1)!!}{r^{2l+1}}\delta_{j<a_l}\nM_{L-1>}\nG_{iK}  \Bigg]
\eea
We can then write down the solution

\be\label{chiijfullsol}
\chi_{ij}^{\mbox{\scriptsize p}}  = 4\sum_{k=0}^\infty\sum_{l=0}^\infty\sum_{q=0}^{[(k+l)/2]} \frac{\left(\delta \hat{x}\right)^q_{KL}}{k!l!} \left[T^{(q)}_{ij KL} - \frac{1}{2}\delta_{ij}T^{(q)}_{mm KL} \right],
\ee
where 

\bea\label{eq:tqijkl}
\fl T^{(q)}_{ij KL} = kl\frac{(2k-1)!!(2l-1)!!}{r^{2k+2l-2q}}c_q(k+l,-(2k+2l+2);r)\delta_{i<a_k}\nM_{K-1>} \delta_{i<a_l}\nM_{L-1>} \nn \\ 
 - k\frac{(2k-1)!!}{r^{2k-2q-1}}c_q(k+l,-(2k+1);r)\delta_{i<a_k}\nM_{K-1>}\nG_{jL} \nn \\
 - l\frac{(2l-1)!!}{r^{2l-2q-1}}c_q(k+l,-(2l+1);r)\delta_{j<a_l}\nM_{L-1>}\nG_{iK} \nn \\
+ r^{2q+2}c_q(k+l,0;r)\nG_{iK}\nG_{jL} .
\eea 
A form of the general solution to Eq.(\ref{fieldeqD}) is thus

\bea
\chi_{ij} &=& 4\sum_{k=0}^\infty\sum_{l=0}^\infty\sum_{q=0}^{[(k+l)/2]} \frac{\left(\delta \hat{x}\right)^q_{KL}}{k!l!} \left[T^{(q)}_{ij KL} - \frac{1}{2}\delta_{ij}T^{(q)}_{mm KL} \right] \nn \\
&& + \sum_{l=0}^\infty\left[ \frac{(-1)^{l+1}}{l!}C_{ijL}\partial_L\frac{1}{r} - \frac{1}{l!}B_{ijL}x^L\right]. 
\eea
Again the moments $C_{ijL}$ and $B_{ijL}$ are partially determined by the gauge condition (\ref{chigauge}). In the case of weakly self-gravitating bodies, the remaining undetermined pieces of the moments parametrizing the homogeneous solution can be expressed in terms of the stress-energy tensor of the system.
In section \ref{sec:energy} below, we will need the trace of the term $C_{ij}$ [cf. Eq.(\ref{chisol}) above] to complete the derivation of the evolution equation for the energy of a body. This term turns out to be fully determined by the gauge condition (\ref{chigauge}). Substituting (\ref{chisol}) into (\ref{chigauge}) yields

\be\label{chigaugeII}
\partial_j\left\{\sum_{l=0}^\infty \left[\frac{(-1)^{l+1}}{l!}C_{ijL}\partial_L\frac{1}{r} - \frac{1}{l!}B_{ijL}x^L\right]\right\} =  -\dot{\zeta}_i -\partial_j\chi_{ij}^{\mbox{\scriptsize p}}.
\ee
From the field equation (\ref{fieldeqD}), we see that $\partial_j\chi_{ij}^{\mbox{\scriptsize p}}$ is harmonic. We can therefore expand it in the usual form

\be\label{divchi}
\partial_j\chi_{ij}^{\mbox{\scriptsize p}} = \sum_{l=0}^\infty \left[\frac{(-1)^{l+1}}{l!}Q_{iL}\partial_L\frac{1}{r} - \frac{1}{l!}\bar{Q}_{iL}x^{<L>}\right], 
\ee
where $Q_{iL}$ and $\bar{Q}_{iL}$ are STF tensors on their multi-index $L$. Substituting this expansion in Eq.(\ref{chigaugeII}) gives

\bea\nn
\fl \sum_{l=0}^{\infty} \left[\frac{(-1)^{l+1}}{l!} C_{ijL}\partial_{jL}\frac{1}{r} - \frac{1}{l!}lB_{ijjL-1}x^{<L-1>}\right]= \sum_{l=0}^\infty \left[\frac{(-1)^l}{l!}lC_{i<L>}\partial_{L}\frac{1}{r} - \frac{1}{l!}B_{ijjL}x^{<L>} \right]\\
= \sum_{l=0}^\infty \left[\frac{(-1)^l}{l!} \Big[\dot{Z}_{iL} + Q_{iL} \Big] \partial_L\frac{1}{r} + \frac{1}{l!}\Big[\dot{Y}_{iL} + \bar{Q}_{iL}\Big]x^{<L>}\right].
\eea
Comparing both sides term by term yields, for $l \geq 1$

\be\label{gaugeresult}
C_{i<L>} = \frac{1}{l}\left(\dot{Z}_{iL} + Q_{iL}\right).
\ee
Hence we get

\be
C_{kk} = Q_{kk},
\ee
since $Z_{jj} $ equals zero from the gauge condition (\ref{adaptedconditionmu}). The problem is thus reduced to identifying the terms in the divergence of the particular solution $\chi^{\mbox{\scriptsize p}}_{ij}$ [cf. Eq.(\ref{chiijfullsol})] that are proportional to $x^k/r^3$. To complete this straightforward albeit slightly tedious calculation, the following formulae, derived using the standard "peeling" identity [e.g. Eq.(4.25) of Ref. \cite{dsxII}], are useful

\be
T_{<L}\delta_{j>j} = \frac{(2l+3)}{(2l+1)}T_L,
\ee
\be
T_{j<L}\delta_{j>i} = \frac{1}{(l+1)(2l+1)}T_{iL},
\ee
\be
T_{<L}\delta_{j>k}T_{<L}\delta_{k>j} = \frac{(2l+3)}{(l+1)(2l+1)^2}T_LT_L,
\ee
where $T_L$ is STF on all its indices. The result is

\be\label{ckk}
C_{kk} = - 8\sum_{l=0}^\infty \frac{1}{l!}\frac{(l+2)}{(2l+3)}\nM_{jL}\nG_{jL}.
\ee

We will now move on to the derivation of the evolution equations for spin and energy. The evolution equation for the spin is simple enough that the material of this section is not needed for that computation. It will be needed in the case of the energy only.

\section{Spin}\label{sec:spin}

The derivation of the evolution equation for the spin $S_i$ presented in this section is inspired by the works of Thorne and Hartle \cite{thornehartle} and Zhang \cite{zhang}. The method is based on the Landau-Lifshitz formulation of the equations of general relativity \cite{landaulifshitz} as

\be\label{Einsteinsequations}
\mathcal{H}^{\mu\alpha\nu\beta}_{\,\,\,\,\,\,\,\,\,\,\,\;,\alpha\beta} = 16\pi \Big[(-g)T^{\mu\nu} + \mathcal{T}^{\mu\nu}\Big].
\ee
The tensor density $\cal{H}^{\mu\alpha\nu\beta}$ is given by

\be\label{LL}
\mathcal{H}^{\mu\alpha\nu\beta} = \mathfrak{g}^{\mu\nu}\mathfrak{g}^{\alpha\beta} - \mathfrak{g}^{\alpha\nu}\mathfrak{g}^{\beta\mu} ,
\ee
where $\mathfrak{g}^{\alpha\beta} = \sqrt{-g}g^{\alpha\beta}$. In Eq.(\ref{Einsteinsequations}), $T^{\mu\nu}$ is the usual stress-energy tensor and the Landau-Lifshitz pseudotensor $\mathcal{T}^{\mu\nu}$ is given by

\bea\nn
\fl \mathcal{T}^{\alpha\beta} = \frac{1}{16\pi}\Bigg[ \mathfrak{g}^{\alpha\beta}_{\;\,,\lambda}\mathfrak{g}^{\lambda\mu}_{\;\,,\mu} - \mathfrak{g}^{\alpha\lambda}_{\;\,,\lambda}\mathfrak{g}^{\beta\mu}_{\;\,,\mu}  + \frac{1}{2}g^{\alpha\beta}g_{\lambda\mu}\mathfrak{g}^{\lambda\nu}_{\;\,,\rho}\mathfrak{g}^{\rho\mu}_{\;\,,\nu} + g_{\lambda\mu}g^{\nu\rho}\mathfrak{g}^{\alpha\lambda}_{\;\,,\nu}\mathfrak{g}^{\beta\mu}_{\;\,,\rho} \\\label{pseudotensor}
\fl -g_{\mu\nu}\left(g^{\alpha\lambda}\mathfrak{g}^{\beta\nu}_{\;\,,\rho}\mathfrak{g}^{\mu\rho}_{\;\,,\lambda} + g^{\beta\lambda}\mathfrak{g}^{\alpha\nu}_{\;\,,\rho}\mathfrak{g}^{\mu\rho}_{\;\,,\lambda}\right) 
 + \frac{1}{8}\left(2g^{\alpha\lambda}g^{\beta\mu} - g^{\alpha\beta}g^{\lambda\mu}\right)\left(2g_{\nu\rho}g_{\sigma\tau} - g_{\rho\sigma}g_{\nu\tau}\right)\mathfrak{g}^{\nu\tau}_{\;\,,\lambda}\mathfrak{g}^{\rho\sigma}_{\;\,,\mu} \Bigg].
\eea
Following Thorne and Hartle \cite{thornehartle}, we first define the quantity

\be\label{angularmomentum}
J^i_\Sigma = \frac{1}{16\pi}\oint_\Sigma \epsilon_{ijk}\left(x^j\mathcal{H}^{k\alpha 0l}_{\,\,\,\,\,\,\,\,\,\,\,\;,\alpha } + \mathcal{H}^{jl0k}\right)d^2\Sigma_l.
\ee
Here we call $J^i_\Sigma$ an ``enclosed angular momentum'' for convenience, although it does not have a geometric, physical meaning. It is introduced merely as a convenient intermediate mathematical tool useful in the derivation of the spin evolution equation. It is a meaningful, physical angular momentum (for stationary systems) only if one takes the surface $\Sigma$ to be a sphere at spatial infinity \cite{mtw}, which we will never do here. In order to derive the evolution equation for the spin of a given body, one must choose the surface $\Sigma$ to be closed and to lie in the buffer region of that body [see Fig.1 of paper I], i.e. the surface $\Sigma$ must be far enough from the compact object so that it lies in the weak, post-Newtonian (or near zone) gravitational field and must exclude all other bodies. Substituting Eq.(\ref{LL}) and the expansion (\ref{gm00})-(\ref{gmij}) into Eq.(\ref{angularmomentum}), we arrive at

\be\label{expandedangularmomentum}
J^i_\Sigma = \frac{\ve^2}{16\pi}\oint_{\Sigma} \epsilon_{ijk} \left(x^j \zeta_{k,m} + \delta_{km}\zeta_j\right)d^2\Sigma_m + O(\ve^4).
\ee

The evolution equation for the spin is obtained in two steps. First, we compute the enclosed angular momentum explicitly from Eq.(\ref{expandedangularmomentum}) and then take its time derivative. Next we use the following formula \cite{thornehartle}, 

\be\label{spinconservation}
\dot{J}^i_\Sigma = - \oint_\Sigma \epsilon_{ijk} \, x^j \mathcal{T}^{kl} \,\, d^2\Sigma_l.
\ee
Note that the derivation of this formula assumes that the volume enclosed by the surface $\Sigma$ on a given time slice has the topolgy of $\R^3$. It excludes, for example, eternal black holes. By computing the surface integral on the right-hand side of (\ref{spinconservation}) explicitly we obtain the evolution equation for the spin of the body.

We start the derivation by rewriting the expression (\ref{adaptedC}) for the gravitomagnetic potential $\zeta_i$ using (\ref{id}) to get

\be\label{newzeta}
\zeta_i = \sum_{l=0}^\infty \left[-\frac{1}{l!}\frac{(2l+1)!!}{(2l+1)} Z_{iL} \frac{x^{<L>}}{r^{2l+1}} - \frac{1}{l!}Y_{iL}x^{<L>}\right].
\ee
Since $Z_{iL}$ is an STF tensor on its $L$ indices, the following relations hold: $Z_{iL}x^{<L>} = Z_{i<L>}x^L = Z_{iL}x^L$. Similar relations hold for $Y_{iL}$. The gradient of the gravitomagnetic potential (\ref{newzeta}) is then given by

\be\label{dnewzeta}
\fl \partial_j\zeta_i = \sum_{l=0}^\infty \left\{-\frac{1}{l!}\frac{(2l+1)!!}{(2l+1)} \left[\frac{l}{r^{l+2}}Z_{ijL-1}n^{L-1}- \frac{2l+1}{r^{l+2}} Z_{iL} n^{jL}\right] - \frac{lr^{l-1}}{l!}Y_{ijL-1}n^{L-1}\right\}.
\ee
We next substitute (\ref{newzeta}) and (\ref{dnewzeta}) into (\ref{expandedangularmomentum}). To perform the integrals, we choose the surface $\Sigma$ to be a coordinate sphere of radius $r$. We will drop the $\Sigma$ subscripts from now on since a specific surface has been chosen. We now have

\bea\nn
\fl J^i = \frac{\ve^2}{16\pi}\oint \epsilon_{ijk} \sum_{l=0}^\infty \frac{1}{l!}\left[\frac{(2l+1)!!}{(2l+1)}\frac{(l+1)}{r^{l+1}}Z_{kL}n^{jL} - lr^lY_{kL}n^{jL} \right. \\\nn
\left. - \frac{(2l+1)!!}{(2l+1)}\frac{1}{r^{l+1}}Z_{jL}n^{kL} - r^lY_{jL}n^{kL}\right] r^2d\Omega  + O(\ve^4) \\\label{almostS}
\fl = \frac{\ve^2r^2}{16\pi}\oint \epsilon_{ijk} \sum_{l=0}^\infty \frac{1}{l!}\left[\frac{(2l+1)!!}{(2l+1)}\frac{(l+2)}{r^{l+1}}Z_{kL}n^{jL} - (l-1)r^lY_{kL} n^{jL}\right]d\Omega+ O(\ve^4).
\eea
The integrals appearing in (\ref{almostS}) are evaluated using the following results, taken from Thorne \cite{thorne}

\be\label{unitvectorintegralA}
\frac{1}{4\pi}\oint\, n^{2L+1} \,d\Omega = 0,
\ee
\be\label{unitvectorintegralB}
\frac{1}{4\pi}\oint\, n^{2L} \,d\Omega = \frac{1}{2l+1}\delta_{(i_1i_2}...\delta_{i_{2l-1}i_{2l})},
\ee
and
\bea\nn
\frac{1}{4\pi}\oint \, T_KS_L n^Kn^Ln^i\,\,d\Omega &=& \frac{(l+1)!}{(2l+3)!!}T_{iL}S_L \;\;\mbox{if}\;\; k=l+1  \\\label{STFintegralB}
&=& 0 \;\;\mbox{if}\;\; |k- l|\neq 1.
\eea
Note that the above equation is true only if $T_L$ and $S_L$ are symmetric trace-free tensors. We then obtain after some algebra

\be
J^i = \ve^2\frac{1}{4}\epsilon_{ijk}Z_{kj} + O(\ve^4).
\ee
Using Eq.(\ref{Zstfreduction}), we finally obtain

\be
J^i = \ve^2 S_i + O(\ve^4).
\ee
Equation (\ref{spinconservation}) then allows us to write the spin evolution equation as

\be\label{precessionA}
\dot{S}_i = - \frac{1}{\ve^2}\oint \epsilon_{ijk} x^j \mathcal{T}^{km} r^2 n_m d\Omega  + O(\ve^2).
\ee
The space-space piece of the Landau-Lifshitz pseudo-tensor that enters Eq.(\ref{precessionA}) is

\be\label{spacespace}
\mathcal{T}^{km} = \frac{\ve^2}{4\pi}\left(\partial_k\Phi\partial_m\Phi - \frac{1}{2}\delta_{km}\partial_j\Phi\partial_j\Phi \right) + O(\ve^4).
\ee
The gradient of the Newtonian potential (\ref{adaptedA}) can be written as 

\be\label{gradphi}
\partial_i \Phi = \sum_{l=0}^\infty \frac{1}{l!}\left(V^{\Phi}_Ln^i - \bar{V}^{\Phi}_{iL}\right)n^L,
\ee
where

\bea\label{V}
V^{\Phi}_L &=& \frac{(2l+1)!!}{r^{l+2}}\nM_L
\eea
and
\bea\label{barV}
\bar{V}^{\Phi}_{iL} &=& \frac{(2l+1)!!}{r^{l+3}}\nM_{iL} + r^l\nG_{iL}.
\eea
Substituting (\ref{gradphi}) into (\ref{spacespace}) and then putting (\ref{spacespace}) into (\ref{precessionA}), we obtain, dropping terms of $O(\ve^2)$,

\bea\nn
 \dot{S}_i = -r^3 \epsilon_{ijk} \sum_{p=0}^\infty\sum_{q=0}^\infty \frac{1}{p!q!}\oint \Bigg[\frac{1}{2}\left(V^\Phi_PV^\Phi_Q - \bar{V}^\Phi_{sP}\bar{V}^\Phi_{sQ}\right)n^jn^kn^Pn^Q  \\\nn + \bar{V}^\Phi_{kP}\left(q\bar{V}^\Phi_Q - V^\Phi_Q\right)n^jn^Pn^Q\Bigg] d\Omega \\\nn
 = -r^3 \epsilon_{ijk} \sum_{p=0}^\infty\sum_{q=0}^\infty \frac{1}{p!q!}\oint \left[\bar{V}^\Phi_{kP}\left(q\bar{V}^\Phi_Q - V^\Phi_Q\right)n^jn^Pn^Q\right] d\Omega \\\nn
 = -r^3\epsilon_{ijk}\left\{\sum_{q=0}^\infty \frac{1}{q!(q+1)!}\frac{(q+1)!}{(2q+3)!!}\bar{V}^\Phi_{kjQ}\left(q\bar{V}^\Phi_Q - V^\Phi_Q\right) \right. \\\nn
 \left. + \sum_{p=0}^\infty \frac{1}{p!(p+1)!}\frac{(p+1)!}{(2p+3)!!}\bar{V}^\Phi_{kP}\left[(p+1)\bar{V}^\Phi_{jP} - V^\Phi_{jP}\right]\right\} \\\label{intermediatespin}
 = -r^3\epsilon_{ijk}\sum_{l=0}^\infty \frac{1}{l!(2l+3)!!}\bar{V}^\Phi_{kL}\left[(l+1)\bar{V}^\Phi_{jL} - V^\Phi_{jL}\right].
\eea
The sum over $q$ in the second to last line above does not contribute since $\bar{V}^\Phi_{kjQ}$ is symmetric in $j$ and $k$. Substituting in (\ref{intermediatespin}) the expressions (\ref{V}) and (\ref{barV}) for $V^\Phi_{iL}$ and $\bar{V}^\Phi_{iL}$ respectively, we get

\bea\nn
\fl \dot{S}_i = -r^3\epsilon_{ijk}\sum_{l=0}^\infty \frac{1}{l!(2l+3)!!}\left[-\left(\frac{(2l+1)!!}{r^{l+3}}\right)^2(l+2) \nM_{kL}\nM_{jL} + (l+1)r^{2L}\nG_{jL}\nG_{kL} \right.\\\nn
\lo+\left.  r^l\left(\frac{(2l+1)!!}{r^{l+3}}\right)(l+1)\nM_{kL}\nG_{jL} - r^l\left(\frac{(2l+1)!!}{r^{l+3}}\right)(l+2)\nM_{jL}\nG_{kL}\right] \\\nn
\fl = -r^3\epsilon_{ijk}\sum_{l=0}^\infty \frac{1}{l!(2l+3)!!}\left[-\left(\frac{(2l+1)!!}{r^{l+3}}\right)^2(l+2) \nM_{kL}\nM_{jL} + (l+1)r^{2L}\nG_{jL}\nG_{kL} \right. \\\label{almostP}
 \left. - r^l\left(\frac{(2l+1)!!}{r^{l+3}}\right)(2l+3)\nM_{jL}\nG_{kL} \right] .
\eea
The first two terms inside the brackets of Eq.(\ref{almostP}) do not contribute since they are symmetric in $j$ and $k$. We are left with

\bea\nn
\dot{S}_i &=&  \sum_{l=0}^\infty \frac{r^3 \epsilon_{ijk}}{l!(2l+1)!!}\left[r^l\frac{(2l+1)!!}{r^{l+3}}\nM_{jL}\nG_{kL} \right] \\\label{sdot}
&=& \sum_{l=0}^\infty \frac{1}{l!} \epsilon_{ijk}\nM_{jL}\nG_{kL},
\eea
which is DSX's evolution equation for the spin of a given body [Eq.(4.21c) of \cite{dsxII}]. Note that the $l=0$ term of the sum in (\ref{sdot}) does not contribute since $\nM_i = 0$ in an adapted coordinate system. We can use the expression (\ref{tidalG}) for the tidal moments $\nG_{L}$ in terms of the mass multipole moments of the other bodies of the system to get, restoring body labels,

\be\label{sdotexplicit}
\dot{S}^A_i = \epsilon_{ijm} \sum_{B\neq A}\sum_{k=0}^\infty\sum_{l=1}^\infty \frac{(-1)^k}{k!l!}(2l+2k+1)!!\nM^A_{jL}\nM^B_{K}\frac{n^{BA}_{<mKL>}}{r_{BA}^{k+l+2}}.
\ee
For a system of Newtonian bodies, the evolution equation (\ref{sdot}) for the spin of a given object can be computed directly by using the Newtonian stress-energy conservation equation, which is essentially how DSX derived it. For such bodies, the quantity $S^A_i$ reduces to the usual Newtonian total angular momentum of body $A$ \cite{dsxII}, i.e.

\be\label{nam}
S^A_i = \int_A d^3x \, \rho\, \epsilon_{ijk}x_jv_k,
\ee
where $\rho$ is the rest energy density and $v^i$ is a Cartesian component of the fluid velocity, all quantities being expressed in the adapted coordinate system of that body. One can therefore  say that Eq.(\ref{sdotexplicit}) extends the validity of the classical precession of equinoxes [in the sense that the precession is caused by a coupling of mass multipole moments to an external tidal field and that the coupling has the same functional form as in Newtonian physics] to a large class of compact objects. It is important to remember that (\ref{sdotexplicit}) is a relation between multipole moments defined as parameters of the metric and that for generic compact objects, it is not possible to express these moments in terms of volume integrals over the matter distribution such as (\ref{nam}). Note also that the computations of this paper are not carried out to high enough order to reproduce the standard post-Newtonian spin-spin and spin-orbit precession effects. A derivation of these effects from a surface integral method requires a more accurate definition of spin, again as a parameter of the metric, than the one used in this paper. One requires the metric up to post-2-Newtonian order to define the spin of the body in order to display these effects in our framework. A systematic analysis of this problem requires a generalization of the theory of reference frames to 2PN order and is thus beyond the scope of this paper. However if such a framework were available, the method used here to derive the precession of equinoxes for compact objects should yield the standard spin-spin and spin-orbit precession effects, as well as new terms involving mass and current mulitpole moments. This will be the topic of a further paper. 

Lastly let us recall that in paper I, another set of mass multipole moments ${\cal M}^A_L$ has been introduced when writing down the explicit equations of motion for the worldline of a given body. They are basically mass moments defined in a coordinate system that does not rotate with respect to distant stars. However the fractional difference between ${\cal M}^A_L$ and $\nM^A_L$ is of order $\ve^2$ so $\nM^A_L$ or ${\cal M}^A_L$ can be used interchangeably in (\ref{sdotexplicit}).

\section{Energy}\label{sec:energy}

In this section we derive the evolution equation for the energy (or mass; we use these words equivalently in this paper) $\nM + \ve^2\pnM$ of a given body of the system. As explained in section III.B of paper I, the Newtonian contribution $\nM$ to the mass is a constant, so the evolution equation will involve only the post-1-Newtonian contribution $\pnM$. It is important to be clear about what is meant here by "Newtonian mass" and "post-1-Newtonian mass". These masses are defined to be the monopole terms of the homogeneous solutions for the potentials $\Phi$ and $\psi$ respectively [cf. Eqs.(\ref{adaptedA}) and (\ref{adaptedC})]. The name "post-1-Newtonian" mass for the parameter $\pnM$ comes from the fact that $\psi$ first appears in the metric at post-1-Newtonian order. More details are provided below.

The derivation of the energy evolution equation follows a similar path to the derivation of the spin evolution equation presented in the previous section. First, we define an enclosed energy as follows \cite{thornehartle}

\be\label{enclosedenergy}
E_\Sigma = \frac{1}{16\pi}\oint_\Sigma \mathcal{H}^{0\alpha 0j}_{\,\,\,\,\,\,\,\,\,\,\,,\alpha} d^2\Sigma_j.
\ee
As was the case for $J^i_\Sigma$, the quantity $E_\Sigma$ is only a mathematical tool. We choose to call it ``enclosed energy'' for convenience. We do not attribute to it a physical meaning. It represents a physical energy only if one takes $\Sigma$ to be a sphere at spatial or null infinity \cite{mtw}, which we will never do here. The conservation law that contains the information on the evolution of the enclosed energy is \cite{thornehartle}

\be\label{massconservation}
\dot{E}_\Sigma = -\oint_\Sigma \mathcal{T}^{0j}\,d^2\Sigma_j.
\ee
Substituting the definition (\ref{LL}) and the expansion (\ref{gm00})-(\ref{gmij}) into (\ref{enclosedenergy}), we obtain

\be
\fl E_\Sigma = \frac{1}{16\pi}\oint_\Sigma \bigg\{\ve^2 4\partial_j\Phi  + \ve ^4 \dot{\zeta}_j +  \ve^4\partial_j\left(4\psi - 8\Phi^2 - \chi_{kk}\right)\bigg\}d^2\Sigma_j + O(\ve^6). 
\ee
This expression involves the post-2-Newtonian field $\chi_{ij}$, whose expansion in terms of STF moments has been discussed in section \ref{sec:inverselaplacian}. We again take $\Sigma$ to be a coordinate sphere of radius $r$. This allows us to simplify some terms in the calculation of the enclosed energy by using 

\be
\oint \partial_j f(rn^i) d^2\Sigma_j = r^2\frac{\partial}{\partial r} \oint f(rn^i) d\Omega,
\ee
where $f$ is any function and $n^i = x^i/r$ as usual. We then have, dropping $\Sigma$ subscripts,

\be\label{energyII}
\fl E = \frac{r^2}{16\pi}\frac{\partial}{\partial r} \oint \left[\ve^2 4\Phi + \ve^4\left(4\psi - 8\Phi^2 - \chi_{kk} \right)\right]\,\, d\Omega + \ve^4\frac{r^2}{16\pi} \oint n_j\dot{\zeta}_j \,\,d\Omega. 
\ee
The first two integrals have similar structure and we will therefore consider them together as follows

\be
I_1 = 4\oint \left(\ve^2\Phi + \ve^4\psi\right)\,\,d\Omega.
\ee
From Eqs.(\ref{unitvectorintegralA}) and (\ref{unitvectorintegralB}), we see that only $l=0$ terms of the expansions (\ref{adaptedA}) and (\ref{adaptedB}) can contribute to $I_1$. Since $\nddotM = 0$, $\nG = 0$ and $\pnG = 0$, the surface integral $I_1$ is equal to

\be\label{contribA}
I_1 = - \frac{16\pi}{r}\left(\ve^2\nM + \ve^4\pnM\right).
\ee
The next surface integral in Eq.(\ref{energyII}) is

\be\label{contribB}
I_2 = -\ve^4 8\oint \Phi^2 \,\,d\Omega.
\ee
The square of the Newtonian potential is

\bea\nn
\fl \Phi^2 = \sum_{k,l=0}^\infty\frac{n^Kn^L}{k!l!}\left\{(2k-1)!!\left[\frac{(2l-1)!!}{r^{k+l+2}}\nM_K\nM_L  + 2 \frac{r^l}{r^{k+1}}\nM_K\nG_L\right] + \nG_K\nG_L r^{k+l}\right\}.\\
\eea
From Eqs.(\ref{unitvectorintegralA}) and (\ref{unitvectorintegralB}), only the $k=l$ terms survive the surface integration. The result is

\bea\nn
\fl I_2 = - 32\pi \ve^4 \sum_{l=0}^\infty\frac{1}{l!}\left[\frac{(2l+1)!!}{(2l+1)^2r^{2l+2}}\nM_L\nM_L  + \frac{2}{(2l+1)r}\nM_L\nG_L + \frac{r^{2l}}{(2l+1)!!}\nG_L\nG_L\right]. \\
\eea
We next skip to the integral in (\ref{energyII}) involving the gravitomagnetic potential

\be
I_3 = \ve^4 \oint n_j \dot{\zeta}_j \,\,d\Omega.
\ee
Only $l=1$ terms of the expansion (\ref{adaptedC}) contribute, giving

\be\label{contribC}
I_3 =  -\frac{4\pi}{3} \ve^4 \left(\frac{1}{r^2}\dot{Z}_{jj} + \dot{Y}_{jj}r\right) = 0,
\ee
as both terms vanish by our choice of coordinates [cf. Eqs.(\ref{adaptedconditionA})-(\ref{adaptedconditionB})]. Had they not been eliminated by a coordinate choice, these terms would have been canceled by appropriate contributions in $I_1$ [coming from the gauge moments  $\mu$ and $\nu$] that would have appeared due to a different choice of coordinates. The last remaining surface integral in Eq.(\ref{energyII}) is 

\be
I_4 = -\ve^4 \oint \chi_{kk} \,\, d\Omega.
\ee
Substituting Eq.(\ref{chisol}) in the above yields

\be\label{contribD}
I_4 = 4\pi \ve^4 \left(\frac{1}{r}C_{kk} + B_{kk}\right)  - \ve^4 \oint \chi_{kk}^{\mbox{\scriptsize p}} \,\,d\Omega.
\ee
From Eq.(\ref{chiijfullsol}), the trace of $\chi_{ij}^{\mbox{\scriptsize p}}$ is

\bea
\chi_{kk}^{\mbox{\scriptsize p}} = -2\sum_{k=0}^\infty\sum_{l=0}^\infty\sum_{q=0}^{[(k+l)/2]} \frac{\left(\delta \hat{x}\right)^q_{KL}}{k!l!}T^{(q)}_{jj KL}  \nn \\
= -2\sum_{k=0}^\infty\sum_{l=0}^\infty\sum_{q=0}^{[(k+l)/2]} \frac{2^q (k+l-2q)!}{(k+l)!(k-q)!(l-q)!} T^{(q)}_{jj Q<K-Q \, L-Q>Q}x^{K-Q}x^{L-Q}. \nn \\
\eea
In the integral of $\chi_{kk}^{\mbox{\scriptsize p}}$ over the unit sphere, the only terms that can contribute must have $q=k=l$, otherwise the integral vanishes due to orthogonality of $\{n^{<L>} \}$ or the fact that $T^{(q)}_{jj KL}$ is STF on $K$ and $L$ separately. We then have

\be\label{eq:intchikk}
\oint \chi_{kk}^{\mbox{\scriptsize p}}  \, d\Omega = -8\pi \sum_{l=0}^\infty \frac{2^l}{(2l)!}T^{(l)}_{jj LL}.
\ee
Substituting (\ref{eq:tqijkl}) into (\ref{eq:intchikk}) yields after some algebra

\bea
\oint \chi_{kk}^{\mbox{\scriptsize p}} \, d\Omega &=& -\frac{4\pi}{r^2}\nM^2 - 8\pi \sum_{l=0}^\infty \frac{1}{l!} \frac{(2l+1)!! }{2(l+1)(2l+3)r^{2l+4}}\nM_{jL}\nM_{jL} \nn \\ \label{chiintfinal}
&& - 8\pi \sum_{l=0}^\infty \frac{1}{l!} \frac{r^{2l+2}}{(2l+2)(2l+3)!!} \nG_{jL}\nG_{jL}. 
\eea
Putting (\ref{ckk}) and (\ref{chiintfinal}) into Eq.(\ref{contribD}), adding together Eqs.(\ref{contribA}), (\ref{contribB}) and (\ref{contribD}) and then substituting the result into Eq.(\ref{energyII}) finally yields

\bea\nn
\fl E = \ve^4 \sum_{l=0}^\infty \frac{1}{l!}\Bigg[\frac{7(l+2)(2l+1)!!}{2(l+1)(2l+3)}\frac{1}{r^{2l+3}}\nM_{jL}\nM_{jL}  + \frac{(2l+3)^2 + 7}{2(l+1)(2l+3)}\nM_{jL}\nG_{jL} \\\label{enclosedmassfinal}
 - \frac{7}{2(2l+3)!!}r^{2l+3}\nG_{jL}\nG_{jL}\Bigg]  + \ve^2 \nM + \ve^4\pnM + \frac{7}{2r}\ve^4 \nM^2.
\eea
We can now make direct use of (\ref{massconservation}). Taking a time derivative of Eq.(\ref{enclosedmassfinal}) gives on one hand

\bea\nn
\fl \dot{E} = \ve^4 \sum_{l=0}^\infty \frac{1}{l!}\Bigg[\frac{7(l+2)(2l+1)!!}{(l+1)(2l+3)}\frac{1}{r^{2l+3}}\ndotM_{jL}\nM_{jL} - \frac{7}{(2l+3)!!}r^{2l+3}\ndotG_{jL}\nG_{jL}   \\\label{mdot}
+ \frac{(2l+3)^2 + 7}{2(l+1)(2l+3)}\big(\ndotM_{jL}\nG_{jL} + \nM_{jL}\ndotG_{jL}\big)  \Bigg] + \ve^4 \pndotM .
\eea
From Eqs. (\ref{gm00})-(\ref{gmij}) and (\ref{pseudotensor}), we have

\be\label{tzeroi}
\mathcal{T}^{0i} = \frac{\ve^4}{4\pi}\Big[\partial_j\Phi\left(\partial_i\zeta_j - \partial_j\zeta_i\right) + 3\dot{\Phi}\partial_i\Phi \Big] + O(\ve^6).
\ee
The computation of the surface integral in the right-hand side of (\ref{massconservation}) using (\ref{tzeroi}) is performed with the same methods as used before. There is nothing new in this computation and therefore we will not report the details here. The result is

\bea\nn
\oint \mathcal{T}^{0i} d^2\Sigma_i = \ve^4 \sum_{l=0}^\infty \frac{1}{l!(2l+3)}\Bigg[\frac{7r^{2l+3}}{(2l+1)!!}\ndotG_{jL}\nG_L  - \frac{(l+5)}{(l+1)}\ndotM_{jL}\nG_{jL} \\\label{tzeroiint}  
+ \frac{(l-2)}{(l+1)}\nM_{jL}\ndotG_{jL} - \frac{7(l+2)(2l+1)!!}{(l+1)r^{2l+3}}\ndotM_{jL}\nM_{jL}  \Bigg]. 
\eea
The conservation law (\ref{massconservation}) then yields

\bea\nn
\pndotM^A = \sum_{l=0}^\infty \frac{1}{l!}\Bigg\{\left[\frac{(l+5)}{(l+1)(2l+3)} - \frac{(2l+3)^2 + 7}{2(l+1)(2l+3)}\right]\ndotM_{jL}\nG_{jL} \\\nn
- \left[\frac{(l-2)}{(l+1)(2l+3)} + \frac{(2l+3)^2 + 7}{2(l+1)(2l+3)}\right]\nM_{jL}\ndotG_{jL}\Bigg\} \\\label{energyevolution}
= -\sum_{l=0}^\infty\frac{1}{l!}\Bigg[\ndotM_{jL}\nG_{jL} + \frac{(l+2)}{(l+1)}\nM_{jL}\ndotG_{jL}\Bigg],
\eea
which is equivalent to Eq.(4.21a) of \cite{dsxII} in an adapted coordinate system. As we have done with the spin evolution equation, we can rewrite Eq.(\ref{energyevolution}) in terms of mass multipole moments alone. This form of the evolution equation is, with body labels $A$ restored,

\bea\nn
\fl \pndotM^A = - \sum_{B\neq A}\sum_{k=0}^\infty\sum_{l=1}^\infty \frac{(-1)^k}{k!l!}(2l+2k+1)!!\Bigg[\ndotM^A_{jL}\nM^B_K\frac{n^{BA}_{<jKL>}}{r_{BA}^{k+l+2}}  \\\label{mdotexplicit}
\fl + \frac{(l+2)}{(l+1)}\nM^A_{jL}\ndotM^B_{K}\frac{n^{BA}_{<jKL>}}{r_{BA}^{k+l+2}} 
 - \frac{(l+2)}{(l+1)}(2k+2l+3)\nM^A_{jL}\nM^B_{K}v^{BA}_m\frac{n^{BA}_{<jmKL>}}{r_{BA}^{k+l+3}} \Bigg].
\eea

To gain some physical insight into the meaning of Eq.(\ref{energyevolution}), it is useful to look at the volume integral expression for the mass monopole of, say, body $A$, i.e. $\nM^A + \ve^2\pnM^A$, in the weakly self-gravitating limit. This is \cite{dsxII}

\be\label{pnmass}
\nM^A + \ve^2\pnM^A = \int_A d^3x \left( {\,}^{\mbox{\tiny n}}\!{T^{00}} +
  \varepsilon^2 \,^{\mbox{\tiny pn}}\!{T^{00}} \right)\left(1 + \ve^2 \bm{x}\cdot\bm{\nabla}\Phi\right),
\ee
where ${\,}^{\mbox{\tiny n}}\!{T^{00}}$ and $\,^{\mbox{\tiny pn}}\!{T^{00}}$ are the Newtonian and post-1-Newtonian pieces of the time-time component of the stress-energy tensor of the body in the adapted coordinate system. For a perfect fluid, the Newtonian term $\,^{\mbox{\tiny n}}\!{T^{00}}$ is simply the locally measured rest energy density $\rho$ and the post-1-Newtonian term $\,^{\mbox{\tiny pn}}\!{T^{00}}$ is equal to $\rho (\bm{v}^2 - 2\Phi)$ \cite{weinberg}. The 1PN mass thus includes contributions from internal fluid motion and from the gravitational potential. Note that for systems of interacting bodies, it is the full Newtonian potential that enters (\ref{pnmass}) and so the other bodies directly contribute to the integral expression for the mass of body $A$. One can check, using 1PN hydrodynamical equations of motion, that for an isolated system the mass (\ref{pnmass}) is conserved \cite{dsxII}. For systems of interacting bodies, the tidal terms in the Newtonian potential due to the presence of other bodies are entirely responsible for the time evolution of the 1PN mass. Although expression (\ref{pnmass}) is only valid for a weakly self-gravitating body, one should expect physical contributions to the 1PN mass of compact objects to include internal fluid motion, gravitational binding and the presence of other bodies. However one must rely on numerical computations to study quantitatively the dependence of the 1PN mass on the internal structure and the environment of the body since no formula like (\ref{pnmass}) exists for objects with strong internal gravity.  

Similarly to the spin case, the computations presented here are not performed at high enough order to display spin interactions in the evolution of the mass. One needs a definition of the mass of a body accurate to 2PN order and derive the evolution equation for this 2PN mass to see such effects. A systematic solution to this problem in our framework requires an extension of the theory of reference frames to 2PN order. 
 
As discussed at the end of section \ref{sec:spin}, the moments $\mathcal{M}^A_L$ can be used interchangeably with the moments $\nM^A_L$ in the right-hand side of Eq.(\ref{mdotexplicit}), since the changes produced by such a replacement are fractional corrections of order $\ve^2$. Using the definition of $\mathcal{M}^A_L$ given in Eq.(5.39) of paper I, the quantity $\ndotM^A + \ve^2\pndotM^A$ is equal to $\dot{\mathcal{M}}^A$. An equivalent form for the evolution of the energy, setting now $\ve = 1$, is thus\footnote{The right-hand side of this equation was denoted $\mathcal{F}\left[\cmz^B_i,\cmdotz^B_i,\mathcal{M}^B_L,\dot{\mathcal{M}}^B_L\right]$ in Eq.(1.7b) of paper I.}

\bea\nn
\fl\dot{\mathcal{M}}^A = -\sum_{B\neq A}\sum_{k=0}^\infty\sum_{l=1}^\infty \frac{(-1)^k}{k!l!}(2l+2k+1)!! \Bigg[\dot{\mathcal{M}}^A_{jL}\mathcal{M}^B_K\frac{n^{BA}_{<jKL>}}{r_{BA}^{k+l+2}} + \frac{(l+2)}{(l+1)}\mathcal{M}^A_{jL}\dot{\mathcal{M}}^B_{K}\frac{n^{BA}_{<jKL>}}{r_{BA}^{k+l+2}} \\
 - \frac{(l+2)}{(l+1)}(2k+2l+3)\mathcal{M}^A_{jL}\mathcal{M}^B_{K}v^{BA}_m\frac{n^{BA}_{<jmKL>}}{r_{BA}^{k+l+3}} \Bigg].
\eea

\section{Conclusion}
In this paper, we derived, using a surface integral method, leading order evolution equations for the spin $S^A_i$ and the energy $\nM^A + \ve^2\pnM^A$ of an astrophysical body part of a multibody system. They coincide with the results of DSX in the limit of weakly self-gravitating bodies. The derivation is valid for a wide class of compact objects and takes into account all mass and current multipole moments. These evolution equations had been merely assumed in paper I to obtain the final form of the translational equations of motion. This paper completes the results of paper I.  As part of the computational method we have constructed an explicit expansion of the general solution to a post-2-Newtonian vacuum field equation in a region bounded by two spherical shells. This type of expansion of post-2-Newtonian fields can in principle be used derive the equations of motion given in paper I without invoking results derived for globally weak fields, following the lines of the computation presented in section \ref{sec:energy}. Future work using our expansions of post-2-Newtonian fields could involve, for example, an extension to post-2-Newtonian order of the theory of reference frames as presented in paper I, studies of gauge invariance of tidal heating and conserved quantities at post-2-Newtonian order. A conserved energy to post-2-Newtonian order in terms of the center-of-mass worldlines and post-1-Newtonian mass and current mulitpole moments could prove useful in astrophysical applications like the gravitomagnetic tidal excitations of {\it r}-modes in coalescing neutron star binaries \cite{gmrmode}.

\ack 

The author is grateful to \'{E}anna Flanagan for helpful discussions and advice in the preparation of this paper. The author also wishes to acknowledge invaluable comments and suggestions from an anonymous referee. The author was supported by Le Fonds Qu\'{e}b\'{e}cois de Recherche sur la Nature et les Technologies (NATEQ) and by NSF grant PHY-0140209.


\section*{References}


\begin{thebibliography}{14}
\bibitem{paperI}
\'{E}. Racine and \'{E}.\'{E}. Flanagan, Phys. Rev. D {\bf 71}, 044010 (2005).

\bibitem{eih}
A. Einstein, L. Infeld and B.Hoffmann, Ann. Math. {\bf 39}, 65 (1938).

\bibitem{landaulifshitz}
L.D. Landau and E.M. Lifshitz, {\it The classical theory of fields;
vol. 2, 4th ed.}, Butterworth Heinemann, 1998, section 96.

\bibitem{thornehartle}
K.S. Thorne and J.B. Hartle, Phys. Rev. D {\bf 31}, 1815 (1985).

\bibitem{dsxII}
T. Damour, M. Soffel and C. Xu, Phys. Rev. D {\bf 45}, 1017 (1992).

\bibitem{bdell}
L. Blanchet and T. Damour, Phil. Trans. R. Soc. Lond. A {\bf 320}, 379 (1986).

\bibitem{bd}
L. Blanchet and T. Damour, Ann. Inst. Henri Poincar\'{e} Phys. Theor. {\bf 50}, 377 (1989).

\bibitem{mtw}
C.W. Misner, K.S. Thorne and J.A. Wheeler, {\it Gravitation}, W.H. Freeman and Company, 1973.

\bibitem{thorne}
K.S. Thorne, Rev. Mod. Phys. {\bf 52}, 299 (1980).

\bibitem{dsxI}
T. Damour, M. Soffel and C. Xu, Phys. Rev. D {\bf 43}, 3273 (1991).

\bibitem{dsxIII}
T. Damour, M. Soffel and C. Xu, Phys. Rev. D {\bf 47}, 3124 (1993).

\bibitem{dsxIV}
T. Damour, M. Soffel and C. Xu, Phys. Rev. D {\bf 49}, 618 (1994).

\bibitem{kop}
S.M. Kopeikin, Celest. Mech. {\bf 44}, 87 (1988); AZh {\bf 67}, 10 (1990) [English translation Soviet Astron. {\bf 34}, 5]; Maniscripta Geod. {\bf 16}, 301 (1991).

\bibitem{Brumberg1989}
V. A. Brumberg and V.A. Kopeikin, in {\it Reference frames in
Astronomy and Geophysics}, ed. J. Kovalevsky, I.I. Mueller and
B. Kolaczek, Kluwer, Dordrecht (1989).

\bibitem{Brumberg1991}
V. A. Brumberg, {\it Essential Relativistic Celestial Mechanics},
Hilger, Bristol (1991).

\bibitem{klionervoinov}
S.A. Klioner and A.V. Voinov, Phys. Rev. D {\bf 48}, 1451 (1993).

\bibitem{zhang}
X.H. Zhang, Phys. Rev. D {\bf 31}, 3130 (1985).

\bibitem{weinberg}
S. Weinberg, \textit{Gravitation and Cosmology: principles and
  applications of the general theory of relativity}, John Wiley \&
Sons Inc., 1972.

\bibitem{gmrmode}
\'{E}. \'{E}. Flanagan and \'{E}. Racine, {\it Gravitomagnetic Resonant Excitation of Rossby Modes in Coalescing Neutron Star Binaries} (in preparation).

\end{thebibliography}
\end{document}